\documentclass[journal,comsoc]{IEEEtran}
\usepackage{amsmath,amsfonts}
\usepackage{algpseudocode}
\usepackage{algorithm}
\usepackage{algorithmicx}
\usepackage{subfigure}
\usepackage{array}
\usepackage[caption=false,font=normalsize,labelfont=sf,textfont=sf]{subfig}
\usepackage{textcomp}
\usepackage{stfloats}
\usepackage{url}
\usepackage{verbatim}
\usepackage{graphicx}
\usepackage{amsmath}
\usepackage{amssymb}
\usepackage{cite}
\usepackage{footnote}
\def\BibTeX{{\rm B\kern-.05em{\sc i\kern-.025em b}\kern-.08em
    T\kern-.1667em\lower.7ex\hbox{E}\kern-.125emX}}
\usepackage{balance}

\usepackage{xcolor}
\definecolor{mygreen}{rgb}{0.78, 0.93, 0.8}

\begin{document}
\title{\huge Joint Antenna Position and Transmit Power Optimization for Pinching Antenna-Assisted ISAC Systems }

\author{Yunhui Qin,
Yaru Fu, {\it{Member,  IEEE}},
and
Haijun Zhang, {\it{Fellow,  IEEE}}

\thanks{
This work was supported  in part by the National Natural Science Foundation of China  under Grant 62401060  and in part by the...6.
(\emph{Corresponding author: Yaru Fu}.)

Yunhui Qin and Haijun Zhang are with the National School of Elite Engineering, Beijing Engineering and Technology Research Center for Convergence Networks and Ubiquitous Services, University of Science and Technology Beijing, Beijing, 100083, China (qinyunhui@ustb.edu.cn, haijunzhang@ieee.org).

Yaru Fu is with the School of Science and Technology, Hong Kong Metropolitan University, Hong Kong, China (e-mail: yfu@hkmu.edu.hk).

 }}

\markboth{Journal of \LaTeX\ Class Files,~Vol.~18, No.~9, July~}%
{How to Use the IEEEtran \LaTeX \ Templates}

\maketitle

\begin{abstract}
This letter explores how pinching antennas, an advanced flexible-antenna system,  can enhance the performance of integrated sensing and communication (ISAC) systems by leveraging their adaptability, cost-effectiveness, and ability to facilitate line-of-sight transmission. 
To achieve this, a joint antenna positioning and transmit power optimization problem is formulated to maximize the total communication data rate while meeting the target sensing requirements and the system energy constraint.
To address the complex non-convex optimization problem, we propose a maximum entropy-based reinforcement learning (MERL) solution. 
By maximizing cumulative reward and policy entropy, this approach effectively balances exploration and exploitation to enhance robustness. 
Numerical results demonstrate that the proposed MERL algorithm surpasses other benchmark schemes in cumulative reward, total data rate, sensing signal-to-noise ratio, and stability. 
\end{abstract}

\begin{IEEEkeywords}
Antenna positioning,
integrated sensing and communication,
pinching antenna,
reinforcement learning,
transmit power.
\end{IEEEkeywords}

\section{Introduction}
\IEEEPARstart{I}{ntegrated} sensing and communication (ISAC) has emerged as a groundbreaking technology that plays a pivotal role in shaping future wireless communication systems \cite{zhang2024joint}.
It can incorporate radar sensing and wireless communication functions on a platform that leverages shared resources such as hardware, spectrum, and energy, thus improving operational efficiency, decreasing costs, and promoting sustainability \cite{liu2020joint}.
This capability addresses the extensive application demands of future intelligent scenarios such as autonomous driving, smart cities, the industrial Internet of Things, and other domains \cite{ zhang2023ustb}.

The utilization efficiency of shared resources and the adaptability to dynamic environments present major challenges  for ISAC systems, especially in their antenna systems.
On one hand, 
the sharing of resources between sensing and communication functions necessitates the use of highly directional and flexible antennas to mitigate  signal interference and improve resource  efficiency.
On the other hand,  in dynamic and complex environments, antennas must be capable of dynamically adapting to environmental variations, thus ensuring high communication quality and sensing accuracy.
However, traditional fixed-position antenna designs often struggle to adapt to these diverse requirements.

In recent years, flexible antennas, such as reconfigurable intelligent surfaces \cite{liu2023integrated}, intelligent reflecting surfaces \cite{wu2019intelligent}, fluid antenna systems \cite{wong2020fluid}, and movable antennas \cite{zhu2023modeling}, have attracted considerable attention. Due to their ability to dynamically reconstruct the wireless channel, flexible antenna systems can significantly enhance performance compared to traditional fixed-position antenna systems \cite{Wang2025Antenna}.
However, in most existing flexible antennas, the variation range of antenna positions is generally limited to the wavelength scale, constraining their ability to counteract large-scale path loss.
In addition, the high costs of many existing flexible antennas also restrict their applications in real-world scenarios.
The emergence of pinching antennas overcomes these challenges, as they can create new line-of-sight (LoS) links or enhance existing transceiver channels by using low-cost dielectric materials placed at arbitrary positions on a dielectric waveguide \cite{ding2024flexible}.

Unlike traditional antennas, pinching antennas can be flexibly deployed, and increasing the number of pinching antennas incurs almost no additional cost \cite{ding2024flexible}.
It can be observed that both the pinching antenna positions and  transmission power significantly affect the performance of the wireless systems.
In \cite{Wang2025Antenna}, the authors studied the optimal locations and number of pinching antennas to activate in order to maximize the throughput of a non-orthogonal multiple access-assisted pinching antenna system and proposed a practical and low-complexity solution.
Later,
the authors of \cite{xu2025rate} optimized the locations of the pinching antennas to maximize the downlink transmission rate.
Although the effectiveness of pinching antennas is verified in  \cite{Wang2025Antenna, ding2024flexible, xu2025rate}, to the best of the authors' knowledge, the potential of optimizing pinching antenna positions and power allocation in ISAC systems has not been fully explored,
which stimulate our work. For brevity, the main contributions of this letter are summarized below:
\begin{itemize}
\item We  investigate  pinching antenna-assisted ISAC systems and formulate a joint optimization problem for pinching antenna positions and power allocation to improve communication rate and sensing  signal-to-noise ratio (SNR).
\item We propose a maximum entropy-based reinforcement learning (MERL) algorithm to  solve the total data rate maximization problem while meeting sensing requirements  and the system energy constraint.
By focusing on maximizing cumulative reward and policy entropy, this approach  ensures a proper balance between exploration and exploitation, resulting in greater robustness.
\item The numerical results show that the proposed MERL algorithm significantly outperforms other benchmarks, achieving  higher cumulative  reward, total data rate, sensing SNR, and  stability.
Specifically, the proposed MELR algorithm has improved the normalized total data rate 
by 20.3\% compared to the twin delayed deep deterministic policy gradient (TD3) \cite{fujimoto2018addressing} and by 44.4\% compared to the deep deterministic policy gradient (DDPG) \cite{silver2014deterministic}. Additionally, it has  improved the normalized sensing SNR by 16.7\% over the TD3 and far surpasses that of the DDPG algorithm.
\end{itemize}

The reminder parts of this letter are given as follows. In Section \ref{sec:model}, we elaborate on the considered system model and formulate the total data rate maximization problem  subject to target sensing constraints. In Section \ref{sec:solution}, we reconstruct the  optimization problem and introduce a MERL solution. 
In Section \ref{sec:simu}, extensive numerical results are presented to evaluate
the proposed algorithm compared with benchmarks.
Finally, we conclude this paper in Section \ref{sec:conclusion}.

\section{System Model and Problem Formulation}\label{sec:model}
\begin{figure}[htbp]
  \centering
  \includegraphics [scale=0.5]{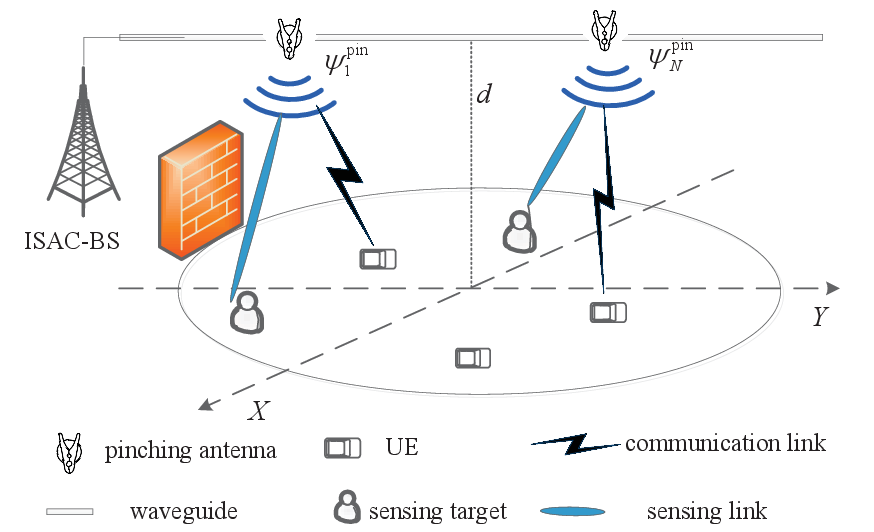}
  \caption{Illustration of  pinching antenna-assisted ISAC system, enabling both UE service and target sensing via LoS links established by pinching antennas. }
  \label{fig:deploy}
\end{figure}

As illustrated in Fig.~\ref{fig:deploy}, we consider a pinching antenna-assisted ISAC system that consists of a base station (BS) equipped with $N$ pinching antennas, serving $M$ single-antenna mobile users (UEs) and $K$ sensing targets. The pinching antennas establish new LoS transceiver links.
Let $\psi_{m,t} = (x_{m,t}, y_{m,t}, 0)$ denote the location of the $m$-th UE in time slot $t$, define $\psi^{\rm{pin}}_{n,t} = (x^{\rm{pin}}_{n,t}, 0, d)$  as the location of the pinching antenna $n$, and denote by $\psi^{\rm{sen}}_k = (x_{k,t}, y_{k,t}, 0)$  the location of the sensing target $k$.
To facilitate system description and algorithm
design, the collections of all pinching antennas, mobile UEs, and
sensing targets are represented by $\mathcal{N} = \{1, 2, \cdots  N\}$,
$\mathcal{M} = \{1, 2, \cdots  M\}$, and $\mathcal{K} = \{1, 2, \cdots  K\}$,  respectively.

Based on the spherical wave channel model, in time slot $t$,  $t \in \mathcal{T} = \{1,2, \cdots T\}$, the chennel vector of $m$-th UE can be expressed as follows \cite{ding2024flexible}: 
\begin{equation}
  {\boldsymbol{h}}_{m,t} = \left[ \frac{\alpha e^{-j\frac{2\pi}{\lambda}{\lvert \psi_{m,t} - \psi^{\rm{pin}}_{1,t} \rvert}}}{\lvert\psi_{m,t} - \psi^{\rm{pin}}_{1,t}\rvert} \cdots \frac{\alpha e^{-j\frac{2\pi}{\lambda}{\lvert\psi_{m,t} - \psi^{\rm{pin}}_{N,t} \rvert}}}{\lvert\psi_{m,t}- \psi^{\rm{pin}}_{N,t} \rvert} \right]^T,
\end{equation}
where $\alpha=\frac{c}{4\pi f_c}$, $c$ denotes the speed of light, $f_c$ is the carrier frequency, and $\lambda$ represents wavelength. 

The $N$ pinching antennas are located on the same waveguide, meaning that the signal transmitted by one pinching  antenna is a phase-shifted version of that transmitted by another. 
Let $x_{c,t}$ denote the information signal passed onto the waveguide, $x_{s,t}$ represent the dedicated sensing signal.
Therefore, the signal vector ${\boldsymbol{x}}$ can be expressed as 
\begin{equation}
  {\boldsymbol{x}} = \left[ \sqrt{\frac{P}{N}}e^{-j\theta_{1,t}}  \cdots \sqrt{\frac{P}{N}}e^{-j\theta_{N,t}}\right]^T (x_{c,t}+x_{s,t}),
\end{equation}
where $P$ is the total transmit power,
$\theta_{n,t}$ is the signal phase shift through antenna $n$ and $\theta_{n,t} = 2\pi
\frac{\lvert\psi^{\rm{pin}}_0 - \psi^{\rm{pin}}_{n,t} \rvert}{\lambda_0} $.
Thereof,
$\psi^{\rm{pin}}_0$ is the location of feed point of waveguide, 
$\lambda_0$ denotes the waveguide wavelength in a dielectric waveguide, i.e., $\lambda_0 = \frac{\lambda}{n_{\rm{neff}}}$,  in which $n_{\rm{neff}}$ is the effective refractive index of a dielectric waveguide \cite{pozar2021microwave},  and $\lambda = \frac{2\pi}{f_c}$.

The received signal of $m$-th UE  can be denoted by $y_{m,t} = {\boldsymbol{h}}^H_{m,t} {\boldsymbol{x}}  + w_{m,t}$, which can be rewritten as follows:
\begin{equation}
y_{m,t} = \left( \sum_{n=1}^{N}  \frac{\alpha e^{-j\frac{2\pi}{\lambda}{\lvert\psi_{m,t} - \psi^{\rm{pin}}_{n,t} \rvert}}}{\lvert\psi_{m,t} - \psi^{\rm{pin}}_{n,t}\rvert} e^{-j\theta_{n,t}} \right)\sqrt{\frac{P}{N}}(x_{c,t}+x_{s,t})  + w_{m,t},
\end{equation}
where $w_{m,t}$ is the additive white Gaussian noise.

At user receivers, the reception of information signal $x_{c,t}$ may suffer from the interference caused by sensing signal $x_{s,t}$. Nevertheless, 
since $x_{s,t}$ can be generated offline and is known to the users before transmission,
they can cancel the interference from $x_{s,t}$ in advance to facilitate the decoding of $x_{c,t}$ \cite{song2022joint}.

Besides, a time-division multiple access (TDMA) is considered in this scenario. In time slot $t$,   the data rate of $m$-th UE achieved by $N$ pinching antennas is 
\begin{equation}
R_{m,t} = \frac{1}{M}{\rm{log}}_2 \left( 1 + \left \lvert \sum_{n=1}^{N}  \frac{\alpha e^{-j\frac{2\pi}{\lambda}{\lvert\psi_{m,t} - \psi^{\rm{pin}}_{n,t} \rvert}}}{\lvert\psi_{m,t} - \psi^{\rm{pin}}_{n,t}\rvert} e^{-j\theta_{n,t}} \right\rvert^2 \frac{p_{m,t}}{N\sigma^2}\right),
\end{equation}
where  $p_{m,t}$ represents the transmit power for user $m$, $\sigma^2$ is the noise
power.

Moreover, we assume that the pinching antennas are deployed successively, with spacings no smaller than the minimum guide distance $\delta$, to avoid antenna coupling, that is 
\begin{equation}
 \left\vert x_{\widetilde{n},t}^{\rm{pin}} - x_{n,t}^{\rm{pin}} \right\vert \geq \delta, 
\end{equation}
where $n, \widetilde{n} \in \mathcal{N}, n \neq \widetilde{n}$,  $\delta > 0$.  $x_{n,t}^{\rm{pin}}$ and $x_{\widetilde{n},t}^{\rm{pin}}$ denote the positions of the pinching antennas on the x-axis of the coordinate system.

Each communication symbol transmitted by the pinching antennas can be regarded as a radar pulse. The chennel vector of  the sensing target $k$ can be expressed as 
\begin{equation}
  {\boldsymbol{h}}_{k,t} = \left[ \frac{\alpha e^{-j\frac{2\pi}{\lambda}{\lvert \psi^{\rm{sen}}_{k,t} - \psi^{\rm{pin}}_{1,t} \rvert}}}{\lvert\psi^{\rm{sen}}_{k,t} - \psi^{\rm{pin}}_{1,t}\rvert} \cdots \frac{\alpha e^{-j\frac{2\pi}{\lambda}{\lvert\psi^{\rm{sen}}_{k,t} - \psi^{\rm{pin}}_{N,t} \rvert}}}{\lvert\psi^{\rm{sen}}_{k,t} - \psi^{\rm{pin}}_{N,t} \rvert} \right]^T,
\end{equation}
where $ k \in \mathcal{K}$, $\psi^{\rm{sen}}_{k,t}$ denotes the location of the sensing target $k$ in time slot $t$.
In this paper, we substitute the signal-to-noise (SNR) of the echo signal received by the BS with the SNR received by the target, as the two are positively correlated \cite{yu2024security}. 
Therefore, the SNR of target $k$ in time slot $t$ can be denoted by 
\begin{equation}
  \Gamma_{k,t} = \frac{\left\lvert \sum_{n=1}^{N} \frac{\alpha e^{-j\frac{2\pi}{\lambda}{\lvert\psi^{\rm{sen}}_{k,t} - \psi^{\rm{pin}}_{n,t}}\rvert}}{\lvert\psi^{\rm{sen}}_{k,t} - \psi^{\rm{pin}}_{n,t}\rvert} e^{-j\theta_{n,t}}  \right\rvert ^2 \frac{p_{m,t}}{N} }{\left(\left\lvert \sum_{n=1}^{N} \frac{\alpha e^{-j\frac{2\pi}{\lambda}{\lvert\psi_{m,t} - \psi^{\rm{pin}}_{n,t}}\rvert}}{\lvert\psi_{m,t} - \psi^{\rm{pin}}_{n,t}\rvert} e^{-j\theta_{n,t}}  \right\rvert ^2 \frac{p_{m,t}}{N} + \sigma^2 \right) }.  
  \end{equation}
To ensure the sensing performance, it is necessary to adjust the  pinching antennas' positions and the signal transmit power to ensure that the SNR of target is always greater than the threshold $\Gamma_{\rm{th}}$, that is $\Gamma_{k,t} - \Gamma_{\rm{th}} \geq 0$. 

Additionally,  the energy consumed  must remain within the limits of the maximum energy constraint,  referred to as $E$, which is given by
\begin{equation}
  \sum\limits_{t \in \mathcal{T}}\sum\limits_{m \in \mathcal{M}} p_{m,t} \Delta t \leq E,
\end{equation}
where $\Delta t $ denotes the duration in each time slot $t$.

In this letter,
our goal is to maximize the total data rate by jointly
optimizing  pinching antennas' position ${\bf{\Psi}} = \{(x_{n,t}^{\rm{pin}}, 0, d), n \in \mathcal{N}, t \in \mathcal{T}\} $ and the transmit power of users $\boldsymbol{p} = \{p_{m,t}, {m \in \mathcal{M}, t \in \mathcal{T}} \}$, while satisfying the  target sensing requirements and the system energy constraint. 
Mathematically, the optimization problem can be formulated as 
\begin{equation}\label{eq:obj}
\max\limits_{{\bf{\Psi}, \boldsymbol{p}}} \sum_{t \in \mathcal{T}} \sum_{m \in \mathcal{M}} R_{m,t}
\end{equation}
\begin{align}
  & \Gamma_{k,t} - \Gamma_{\rm{th}} \geq 0, \forall k \in \mathcal{K}, \forall t \in \mathcal{T} \tag{9a}\\
  & 0 \leq p_{m,t} \leq P_0, \forall m \in \mathcal{M}, \forall t \in \mathcal{T} \tag{9b} \\
  &\sum\limits_{t \in \mathcal{T}}\sum\limits_{m \in \mathcal{M}} p_{m,t}\Delta t \leq E \tag{9c}\\
  & \vert x_{\widetilde{n},t}^{\rm{pin}} - x_{n,t}^{\rm{pin}} \vert \geq \delta, \forall n, \widetilde{n} \in \mathcal{N}, n \neq \widetilde{n}, \forall t \in \mathcal{T}\tag{9d} 
  \end{align}
where $P_0$ is the maximum transmit power for each user.
(9a) ensures the SNR requirement of target sensing,
(9b) and (9c) are the constraints of power allocation and consumed energy, respectively.
(9d) guarantees the  constrains of  antenna spacings.
The optimization problem (\ref{eq:obj}) is clearly nonconvex and further complicated by factors such as user mobility, variations in antenna positioning, and dynamic channel conditions, making it challenging to solve directly.

\section{Maximum Entropy-based Reinforcement Learning Solution }\label{sec:solution}
This section reconstructs the optimization problem and introduces the MERL algorithm to address the complex and nonconvex nature of the problem.
\subsection{The Problem Reconstruction}
The optimization objective delineated in (\ref{eq:obj}) presents a mathematically intractable challenge due to its inherently nonlinear and nonconvex characteristics. To address this, we conceptualize the configuration of pinching antenna positions alongside the  transmit power allocation to users as a sequential decision-making issue.
Let $S$ denote the space of states, $A$ denote  the
space of actions, $R$ represent the reward function, and $(S,A,R,\gamma)$  form a tuple representing a decision process, where $\gamma$ is the discount factor.
\begin{itemize}
  \item  {\bf{State space}} $S$: In time slot $t$, the state $s_t \in S$ mainly includes   the  locations of all pinching antennas,  users and the sensing targets,
  along with the total energy,
      \begin{equation}
      \begin{aligned}
        s_t= \left\{ {\bf{\Psi}}_t, {\bf{\Psi}}^{\rm{UE}}_t, {\bf{\Psi}}^{\rm{Target}}_t, E_t \right\},
      \end{aligned}
      \end{equation}
       \noindent where  $ {\bf{\Psi}}_t = \{ \psi^{\rm{pin}}_{1,t},\cdots \psi^{\rm{pin}}_{N,t} \}$,  ${\bf{\Psi}}^{\rm{UE}}_t = \{\psi_{1,t},\cdots \psi_{M,t}\}$ and $ {\bf{\Psi}}^{\rm{Target}}_t = \{ \psi^{\rm{sen}}_{1,t},\cdots \psi^{\rm{sen}}_{K,t} \}$ are the locations of pinching antennas, users and sensing target in time slot $t$, respectively. $E_t$ represents the total energy constraints of the system under consideration. 
  \item {\bf{Action space}} $A$: The action $a_t \in A$ is defined as the variations in pinching antennas' locations, users' locations, and the power allocation for users,
       \begin{equation}
       \begin{split}
        a_t= & \left\{  \Delta \Psi_t, \Delta \Psi^{\rm{UE}}_t, \boldsymbol{p}_t, \Delta E_t \right\},
       \end{split}
       \end{equation}
        \noindent where  $\Delta \Psi_t = {\bf{\Psi}}_t - {\bf{\Psi}}_{t-1}$ represents the  displacements of  pinching antennas, $\Delta \Psi^{\rm{UE}}_t = {\bf{\Psi}}^{\rm{UE}}_t - {\bf{\Psi}}^{\rm{UE}}_{t-1}$ represents the displacements of  users, 
        $\boldsymbol{p}_t = \{p_{1,t}, \cdots p_{M,t}\}$  denotes the  allocated  power in time slot $t$,
        $\Delta E_t$  denotes the energy consumed in one time slot, which is allocated to the mobile users during the training process.
  \item {\bf{Reward function}}:  The reward function  $r_t$,  $r_t \in R$,  cannot be
  defined the same as (\ref{eq:obj}), since its objective as a sequential
  decision-making problem is constrained by the  SNR requirement of target sensing.
  Combining the original optimization objective (\ref{eq:obj}) and  the  SNR requirement, the reward function can be defined as follows:
  \begin{equation}
    r_t = \max\limits_{\forall m \in \mathcal{M}} R_{m,t} + \beta (\Gamma_{k,t} - \Gamma_{\rm{th}}), 
  \end{equation}
  where $\beta$ denotes the weight of this term in the problem.  
\end{itemize}

\subsection{The MERL Solution}
In this subsection, we present MERL as a solution to this problem. The core idea of  MERL is to not only maximize the cumulative reward but also maximize the entropy of the policy when optimizing the strategy, 
which is given by
\begin{equation}
 \mathop{\max}    \mathbb{E} \left[  \sum_{t=1}^T \gamma^{t-1}  [r_t(s_t, a_t) - \rho{\rm{log}}{\pi_{\phi}}(a_t\lvert s_t)] \right],
\end{equation}
\noindent where $\gamma \in (0,1)$ denotes the discount factor, $\rho$ is the temperature parameter, and $\pi_\phi$ denotes policy network, which is discussed in detail below.
This design ensures a balance between exploration and exploitation, thereby enhancing robustness.

In this algorithm, the agent consists of  the following neural networks:
The policy network $\pi_\phi$ generates
action according to the state of agent, which is a stochastic policy network with a parameter vector  $\phi$.
Moreover, there are two Q-networks $Q_{\theta_1}$ and $Q_{\theta_2}$ with network parameter vectors $\theta_1$ and $\theta_2$ to reduce overestimation, as well as one state value function network. During the training process, the Q-networks take the state vector $s_t$ and the action vector $a_t$ as input.
The state-action value function is derived from soft Bellman equation.

As the agent interacts with the environment in each time slot, a new experience tuple 
$(s_t,a_t,r_t,s_{t+1})$ is generated and stored in the replay memory buffer 
$\bf{D}$. 
With the training of the algorithm, the number of tuples stored in the replay memory buffer incrementally grows, eventually reaching a size sufficient for sampling. The optimization of neural networks is facilitated through the extraction of a mini-batch of experience tuples, denoted as $D$, from the  replay memory buffer,
where $D \subset \bf{D}$ and the size of the mini-batch is $\lvert D \rvert$.

The Q networks  can be updated according to
\begin{equation}
        L_Q = \mathbb{E}_{{(s_t,a_t,r_t, s_{t+1}) \sim D}}
        \left[(y_t-Q_{\theta_i}(s_t,a_t))^2\right], 
\end{equation}
where $i=1,2$ and $y_t$ is the target value of  Q-network, which is given by
\begin{equation}
        y_t=r_t+\gamma
        \left(\min\limits_{i=1,2}Q_{\theta^{'}_i}(s_{t+1},a_{t+1})
        -\rho {\rm{log}}_{\pi_\phi}(\tilde{a}_{t+1}\vert s_{t+1}) \right),
\end{equation}
\noindent where 
$\rho$  indicates the relative importance of the reward compared to the entropy term, introduces randomness into the optimal policy, and can be adaptively optimized.
$\tilde{a}_{t+1}$ emphasizes that the next action should be resampled from the policy.

The policy network  is updated according to
\begin{equation}
\begin{aligned}
        L_{\pi} = \mathbb{E}_{{(s_t,a_t,r_t, s_{t+1}) \sim D}} \left[ \min\limits_{i=1,2}Q_{\theta_i}(s_t,a_t)
        -\rho{\rm{log}}{\pi_{\phi}}(a_t\lvert s_t)\right].
\end{aligned}
\end{equation}

Besides, the target Q networks follow the soft update rule $\theta^{'}_i \gets \epsilon \theta_i +(1-\epsilon)\theta^{'}_i ,i=1,2$,
where $\epsilon$ is the soft update parameter.

\section{Simulation Experiments And Analysis}\label{sec:simu}
In this section, we provide numerical simulations to validate the effectiveness of our developed joint antenna position and power allocation scheme.
In the simulations, the communication users and sensing targets are randomly distributed within a 150 m $\times$ 150 m square area,
and the main parameters are summarized in Table \ref{tab1}.

\begin{table}[]
  \caption{Main Simulation Parameters}
  \vspace{3pt}
  \centering
  \begin{tabular}{p{6cm}p{1.5cm}}
      \hline
      Parameter &  Value \\
      \hline
      Height of pinching antennas, $d$ & $3$ m \\
      The noise power, $\sigma^2$ & $-90$ dBm \\
      Carrier frequency, $f_c$ & $28$ GHz \\
      Effective refractive index of  waveguide, $n_{\rm{neff}}$ & $1.4$  \\
      Sensing SNR threshold, $\Gamma_{\rm{th}}$ & $10$ dB \\
      Pinching antenna spacing, $\delta$ & $\frac{\lambda}{2}$  \\
      Number of UEs, $M$ & $6$  \\
      Number of pinching antennas, $N$ & $3$  \\
      Number of sensing targets, $K$ & $1$ \\
      Time slot, $T$ & $100$ \\
      The discount factor, $\gamma$ & $0.97$ \\
      The size of the mini-batch, $\vert D \vert$ & $256$ \\
      The  soft update parameter, $\epsilon$ & $0.01$ \\
      \hline
  \end{tabular}
  \label{tab1}
\end{table}

To evaluate the performance of our proposed algorithm, three benchmark schemes are adopted in the following parts: 
\begin{itemize}
\item DDPG: This scheme is a policy-based and  off-policy reinforcement learning algorithm, utilizing  a deterministic policy with exploration noise \cite{silver2014deterministic}.
\item TD3: TD3 is an enhanced version of DDPG, employing  two independent critic networks to mitigate overestimation and implementing delayed updates for the actor network to ensure stability \cite{fujimoto2018addressing}.
\item Random scheme: This scheme requires random values for the positions of the pinching antennas and the transmit power, each within their respective predefined regions. 
\end{itemize}

\begin{figure}[htbp]
  \centering
  \includegraphics [scale=0.5]{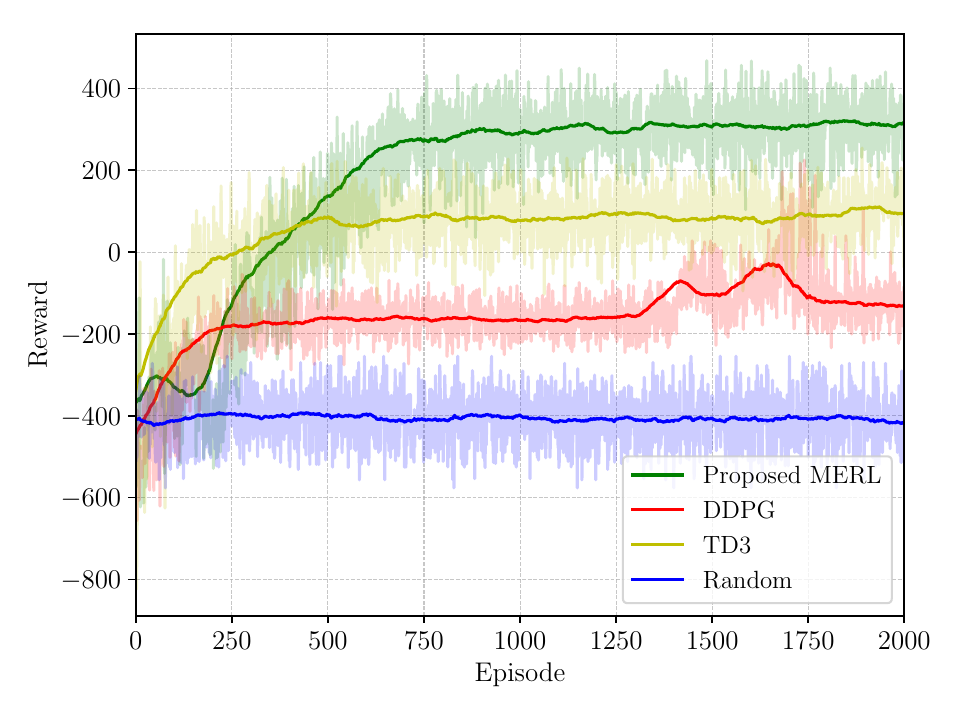}
  \caption{Reward performance of the proposed MELR compared to benchmark schemes. }
  \label{fig:reward}
\end{figure}
Fig.~\ref{fig:reward} shows the reward performance  of the proposed MELR algorithm compared to benchmark schemes, where the learning rate ($LR$) of each RL algorithm is $LR = 1 \times 10^{-5}$.
It can be observed that the proposed MERL algorithm in the considered scenario achieves a higher cumulative rewards in the later stages of algorithm training compared to all benchmarks.
Additionally, the MERL algorithm demonstrates better stability and robustness compared to DDPG scheme, which reveals the superority of maximizing both cumulative reward and policy entropy when optimizing the antenna position and transmit power strategy.

\begin{figure}[htbp]
  \centering
  \subfigure[The normalized  data rate.]{
  \includegraphics[scale=0.5]{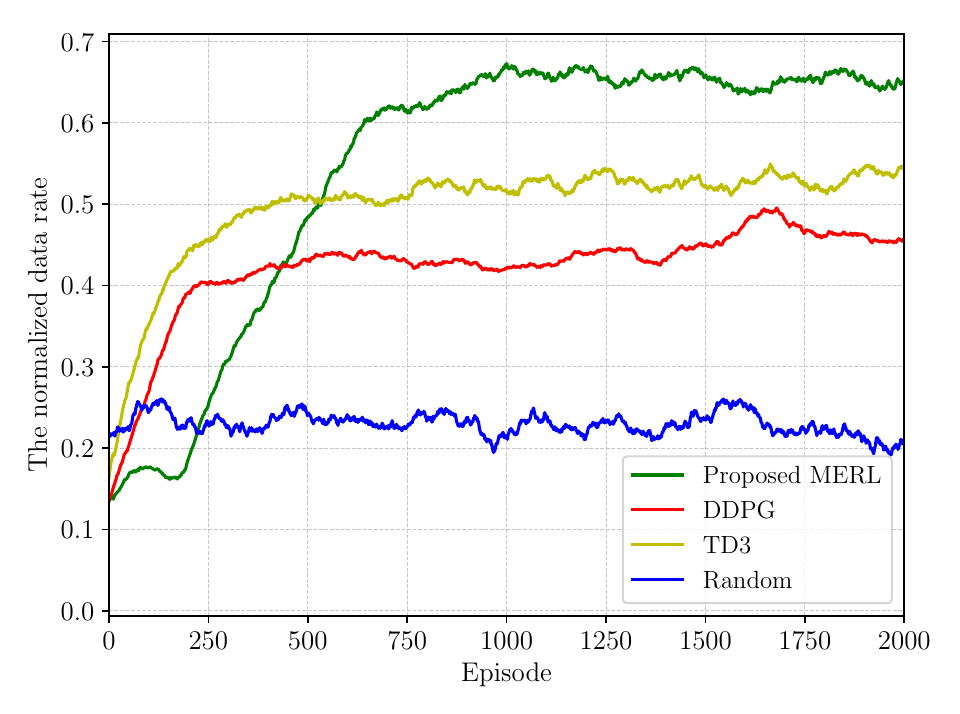}
  }
  \subfigure[The normalized SNR.]{
  \includegraphics[scale=0.5]{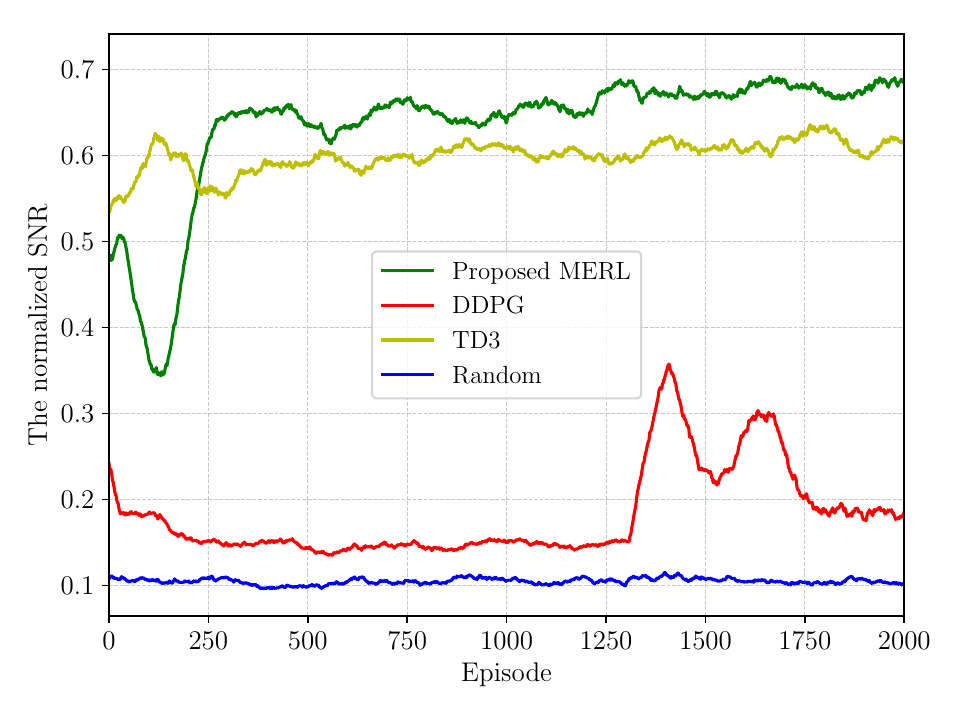}
  }
  \caption{The normalized communication and sensing performance of the proposed MERL algorithm versus benchmarks.}
  \label{fig:rate-snr}
\end{figure}
To facilitate comparison, both the communication and sensing performance are  normalized in our simulations. As shown in  Fig.~\ref{fig:rate-snr}, the normalized total data rate and the normalized sensing SNR for the proposed MERL algorithm and benchmarks are demonstrated.
It can be observed from Fig.~\ref{fig:rate-snr} (a) that the proposed MERL algorithm achieves a higher data rate compared to other benchmarks. Specifically, the proposed
algorithm has improved the normalized total data rate 
by 20.3\% compared to TD3 algorithm and by 44.4\% compared to DDPG algorithm.
Fig.~\ref{fig:rate-snr} (b) shows that the proposed MERL algorithm achieves better normalized sensing SNR and stability compared to other benchmarks. It improves the normalized sensing SNR by 16.7\% over the TD3  and far surpasses that of the DDPG algorithm.

\begin{figure}[htbp]
  \centering
  \includegraphics [scale=0.5]{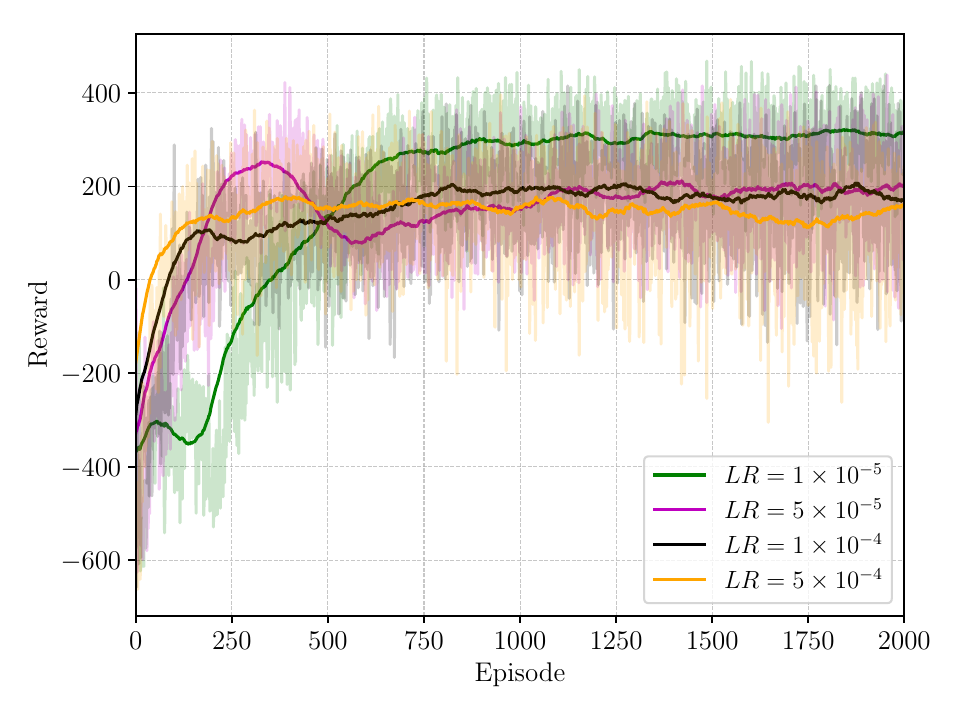}
  \caption{The reward of MELR algorithm versus different learning rates. }
  \label{fig:lr}
\end{figure}
Fig.~\ref{fig:lr} shows the cumulative reward of the MERL algorithm versus different learning rates. 
It can be seen that the proposed MERL algorithm yields the best cumulative rewards in the later stages of algorithm training with a learning rate of $LR= 1 \times 10^{-5}$, and the cumulative reward performance of the MERL algorithm demonstrates relative stability across varying learning rates.

\section{Conclusion}\label{sec:conclusion}
This letter investigated  a pinching antenna-assisted ISAC system by jointly optimizing the pinching antenna positions and user power allocation to enhance both total data rate and sensing SNR. To address the total data rate maximization problem while ensuring the target sensing requirements and the system energy constraint, a MERL solution was introduced. The proposed algorithm maximizes both the cumulative reward and the policy entropy, ensuring a robust strategy for antenna positioning and transmit power allocation that balances exploration and exploitation.
The numerical results demonstrated that the proposed MERL algorithm significantly outperforms other benchmarks, achieving higher cumulative discounted reward, total data rate, sensing SNR and stability.
In future work, we will further investigate the effects of pinching antennas in ISAC networks assisted by non-orthogonal multiple access, as well as the application of off-policy reinforcement learning algorithms in these networks.

\bibliographystyle{IEEEtran}
\bibliography{PA-ISAC-ref}

\begin{thebibliography}{10}
\providecommand{\url}[1]{#1}
\csname url@samestyle\endcsname
\providecommand{\newblock}{\relax}
\providecommand{\bibinfo}[2]{#2}
\providecommand{\BIBentrySTDinterwordspacing}{\spaceskip=0pt\relax}
\providecommand{\BIBentryALTinterwordstretchfactor}{4}
\providecommand{\BIBentryALTinterwordspacing}{\spaceskip=\fontdimen2\font plus
\BIBentryALTinterwordstretchfactor\fontdimen3\font minus
  \fontdimen4\font\relax}
\providecommand{\BIBforeignlanguage}[2]{{%
\expandafter\ifx\csname l@#1\endcsname\relax
\typeout{** WARNING: IEEEtran.bst: No hyphenation pattern has been}%
\typeout{** loaded for the language `#1'. Using the pattern for}%
\typeout{** the default language instead.}%
\else
\language=\csname l@#1\endcsname
\fi
#2}}
\providecommand{\BIBdecl}{\relax}
\BIBdecl

\bibitem{zhang2024joint}
H.~Zhang, B.~Chen, X.~Liu, and C.~Ren, ``Joint radar sensing, location, and
  communication resources optimization in {6G} network,'' \emph{IEEE J. Sel.
  Areas Commun.}, vol.~42, no.~9, pp. 2369--2379, 2024.

\bibitem{liu2020joint}
F.~Liu, C.~Masouros, A.~P. Petropulu, H.~Griffiths, and L.~Hanzo, ``Joint radar
  and communication design: Applications, state-of-the-art, and the road
  ahead,'' \emph{IEEE Trans. Commun.}, vol.~68, no.~6, pp. 3834--3862, 2020.

\bibitem{zhang2023ustb}
H.~Zhang, D.~Wang, S.~Wu, W.~Guan, and X.~Liu, ``{USTB 6G}: Key technologies
  and metaverse applications,'' \emph{IEEE Wireless Commun.}, vol.~30, no.~5,
  pp. 112--119, 2023.

\bibitem{liu2023integrated}
R.~Liu, M.~Li, H.~Luo, Q.~Liu, and A.~L. Swindlehurst, ``Integrated sensing and
  communication with reconfigurable intelligent surfaces: Opportunities,
  applications, and future directions,'' \emph{IEEE Wireless Commun.}, vol.~30,
  no.~1, pp. 50--57, 2023.

\bibitem{wu2019intelligent}
Q.~Wu and R.~Zhang, ``Intelligent reflecting surface enhanced wireless network
  via joint active and passive beamforming,'' \emph{IEEE trans. wireless
  commun.}, vol.~18, no.~11, pp. 5394--5409, 2019.

\bibitem{wong2020fluid}
K.-K. Wong, A.~Shojaeifard, K.-F. Tong, and Y.~Zhang, ``Fluid antenna
  systems,'' \emph{IEEE Trans. Wireless Commun.}, vol.~20, no.~3, pp.
  1950--1962, 2020.

\bibitem{zhu2023modeling}
L.~Zhu, W.~Ma, and R.~Zhang, ``Modeling and performance analysis for movable
  antenna enabled wireless communications,'' \emph{IEEE Trans. Wireless
  Commun.}, vol.~23, no.~6, pp. 6234--6250, 2023.

\bibitem{Wang2025Antenna}
K.~Wang, Z.~Ding, and R.~Schober, ``Antenna activation for {NOMA} assisted
  pinching-antenna systems,'' \emph{IEEE Wireless Commun. Lett.}, pp. 1--1,
  2025.

\bibitem{ding2024flexible}
Z.~Ding, R.~Schober, and H.~V. Poor, ``Flexible-antenna systems: A
  pinching-antenna perspective,'' \emph{arXiv preprint arXiv:2412.02376}, 2024.

\bibitem{xu2025rate}
Y.~Xu, Z.~Ding, and G.~K. Karagiannidis, ``Rate maximization for downlink
  pinching-antenna systems,'' \emph{IEEE Wireless Commun. Lett.}, 2025.

\bibitem{fujimoto2018addressing}
S.~Fujimoto, H.~Hoof, and D.~Meger, ``Addressing function approximation error
  in actor-critic methods,'' in \emph{International conference on machine
  learning}.\hskip 1em plus 0.5em minus 0.4em\relax PMLR, 2018, pp. 1587--1596.

\bibitem{silver2014deterministic}
D.~Silver, G.~Lever, N.~Heess, T.~Degris, D.~Wierstra, and M.~Riedmiller,
  ``Deterministic policy gradient algorithms,'' in \emph{International
  conference on machine learning}.\hskip 1em plus 0.5em minus 0.4em\relax PMLR,
  2014, pp. 387--395.

\bibitem{pozar2021microwave}
D.~M. Pozar, \emph{Microwave engineering: theory and techniques}.\hskip 1em
  plus 0.5em minus 0.4em\relax John wiley \& sons, 2021.

\bibitem{song2022joint}
X.~Song, D.~Zhao, H.~Hua, T.~X. Han, X.~Yang, and J.~Xu, ``Joint transmit and
  reflective beamforming for {IRS}-assisted integrated sensing and
  communication,'' in \emph{Proc. IEEE WCNC}, Austin, TX, USA, May 2022, pp.
  189--194.

\bibitem{yu2024security}
X.~Yu, J.~Xu, N.~Zhao, X.~Wang, and D.~Niyato, ``Security enhancement of {ISAC}
  via {IRS-UAV},'' \emph{IEEE Trans. Wireless Commun.}, vol.~23, no.~10, pp.
  15\,601--15\,612, 2024.

\end{thebibliography}

\end{document}